\documentclass[a4paper,dvips,12pt]{article}
\usepackage{amsmath,amssymb,exscale}   
\usepackage{array,multicol}
\usepackage{afterpage,float,flafter}   
\usepackage{epsfig,rotating,pifont,fancybox}   
\usepackage{cite}
\makeatletter   
\@addtoreset{equation}{chapter}   
\makeatother   
\newcommand{\bea}{\begin{eqnarray}}   
\newcommand{\eea}{\end{eqnarray}}   
   
\newcommand{\NPB}[3]{\emph{ Nucl.~Phys.} \textbf{B#1} (#2) #3}   
\newcommand{\PLB}[3]{\emph{ Phys.~Lett.} \textbf{B#1} (#2) #3}   
\newcommand{\PRD}[3]{\emph{ Phys.~Rev.} \textbf{D#1} (#2) #3}   
\newcommand{\PRL}[3]{\emph{ Phys.~Rev.~Lett.} \textbf{#1} (#2) #3}   
   
\newcommand{\PTP}[3]{\emph{ Prog.~Theor.~Phys.} \textbf{#1}  (#2) #3}   
\newcommand{\IJMP}[3]{\emph{ Int.~J.~Mod.~Phys.} \textbf{#1}  (#2) #3}   
\newcommand{\MPL}[3]{\emph{ Mod.~Phys.~Lett.} \textbf{A#1} (#2) #3}   
\newcommand{\PR}[3]{\emph{ Phys.~Rep.} \textbf{#1} (#2) #3}

\newcommand{\AP}[3]{\emph{ Ann.~Phys.} \textbf{#1} (#2) #3} 
  
\newcommand{\JHEP}[3]{\emph{ JHEP} \textbf{#1} (#2) #3}

\def\simlt{\stackrel{<}{{}_\sim}}

\title{   
\vspace*{-0.8cm}   
\begin{flushright}   
\normalsize{      
IEM-FT-221/01\\
IFT-UAM/CSIC-01-30\\   
\texttt{hep-th/0110132}}\\ 
\end{flushright}    
\vspace{1cm}
\Large{\sc Supersymmetry breaking on orbifolds from\\
Wilson lines~\footnote{Work 
supported in part by CICYT, Spain, under contract AEN98-0816,
and by EU under contracts HPRN-CT-2000-00152 and HPRN-CT-2000-00148.}}
\vspace*{.5cm}
\author{\large
{\sc G.~v.~Gersdorff  and M.~Quir{\'o}s}\\ \\
\emph{Instituto de Estructura de la Materia (CSIC), Serrano 123,}\\
\emph{E-28006-Madrid, Spain.}}}
\date{}   
\begin{document}
\maketitle
\thispagestyle{empty}
\vspace*{.5cm}

\begin{abstract}\noindent
We consider five dimensional theories compactified on the orbifold
$S^1/\mathbb{Z}_2$ and prove that spontaneous local supersymmetry
breaking by Wilson lines and by the Scherk-Schwarz mechanism are
equivalent. Wilson breaking is triggered by the $SU(2)_R$ symmetry
which is gauged in off-shell $N=2$ supergravity by auxiliary
fields. The super-Higgs mechanism disposes of the would-be Goldstinos
which are absorbed by the gravitinos to become massive.  The breaking
survives in the flat limit, where we decouple all gravitational
interactions, and the theory becomes softly broken global
supersymmetry.
\end{abstract}
\vspace{6.cm}   
   
October 2001   
\newpage

\section{\sc Introduction}
Supersymmetry and extra dimensions seem to be essential ingredients in
any theory that aims to solve the hierarchy problem, makes gravity
compatible with quantum mechanics and unifies gravity with the rest of
known particle physics interactions: string theory.  Moreover, the
possibility of lowering the string ($M_s$) and compactification
($1/R$) scales~\cite{lykken,ignas} to values accessible for present
and future accelerators (i.e.~the TeV range) has triggered an enormous
interest in fields theories with large extra dimensions where
gravity~\cite{gravity} and/or matter~\cite{matter} propagate.  In
particular, in theories with compactification scales in the TeV range
where matter propagates the effect of Kaluza-Klein (KK) excitations
can be detected by direct~\cite{direct} or indirect~\cite{indirect}
production at present colliders providing an unambiguous signature of
extra dimensions~\cite{ABrev}. 

These theories present the exciting feature that some observables are
insensitive to the ultraviolet (UV) cutoff
$M_s$~\cite{matter,UV,one,one1,one2}, and thus they are finite and can
be considered as predictions of the theory: this happens in particular
with the Higgs mass in supersymmetric higher dimensional models, a
fact that has opened a strong debate on the subject~\cite{debate}. In
fact summing over all the tower of KK-excitations when they propagate
in internal lines is at the origin of the Higgs mass finiteness in the
previously mentioned models.  It was argued~\cite{debate} that the
presence of a cutoff $\sim M_s$ in the higher dimensional theory
should imply cutting off those sums at KK-modes with masses $M^{(n)}$
such that $M^{(n)}\simlt M_s$, thus generating quadratic divergences
that could not be canceled by supersymmetry. However it was
subsequently argued that regularizing KK-sums with a sharp cutoff
(truncated towers) was not consistent with the symmetries of the
higher dimensional theory.  In particular if the underlying theory is
a string theory (as we are assuming) the couplings of bulk states to
localized states are regularized as
$\exp\{-(M^{(n)}/M_s)^2/2\}$~\cite{strings} and so their interactions
smoothly decouple without spoiling the higher dimensional (tower)
structure. Using the stringy regularization it was shown in
Ref.~\cite{one} that the Higgs mass was finite and coincident with the
previous calculations. Furthermore the same conclusion was reached
using genuine field-theoretical regularization consistent with the
symmetries of the higher dimensional theory, in particular with a
Pauli-Villars regulator~\cite{one2}.  As expected, physical
observables are independent of the regularization procedure provided
the regularization does not spoil any symmetry of the theory.

The ultimate reason for the finiteness of the Higgs mass can be traced back to
the underlying string theory. In type $I/I'$ strings, matter corresponds to
open strings with ends on the brane and thus with KK-modes along  
longitudinal directions and winding modes along transverse directions. 
If transverse
directions are very large (gravitational-like) the winding modes are
extremely heavy and decouple. Moreover in the limit 
$1/R\ll M_s$ (where $R$ is the radius of the longitudinal directions)
string excitations decouple and one recovers the field theory result.
This has been shown in explicit examples~\cite{cuerdas}. 
If the compact dimensions the brane is wrapped around possess the structure
of an orbifold (a possible requirement from chirality) there can be
localized states on the orbifold fixed points which do not feel the compact
dimensions; they are similar to the twisted states of the heterotic string,
required by the anomaly cancellation condition.
On the other hand matter in the bulk for
heterotic strings possesses both KK and winding modes
with respect to compact dimensions and thus the safe procedure is 
considering field theory calculation up to the compactification scale and
include the string result at that scale 
as ``threshold corrections''~\cite{nilleshet}.

The origin of electroweak and supersymmetry breaking is possibly one
of the more poorly known aspects of gauge theories.  In theories with
compactified extra dimensions one can break gauge symmetry or
supersymmetry by imposing different boundary conditions on different
fields (Scherk-Schwarz (SS) mechanism~\cite{SS,SSstring}). In
particular the SS breaking of a gauge symmetry~\cite{SSgauge} can be
seen, if all fields have periodic boundary conditions, as a
spontaneous breaking induced by the vacuum expectation value (VEV) of
the extra dimensional component of a gauge field, which plays the role
of a Higgs boson in four dimensional theories (Wilson line or Hosotani
mechanism~\cite{hosotani}). In this case, if the electroweak Higgs is
associated with a Wilson line the finiteness of the Higgs mass can
follow, even in non supersymmetric string or field
models~\cite{cuerdas,nonsusy}.

On the other hand, the SS-mechanism has been used to break
supersymmetry by imposing non-trivial boundary conditions on fields
transforming non-trivially under a global symmetry of the
theory~\cite{matter,UV}.  Recently SS-supersymmetry breaking based on
the $N=2$ $SU(2)_R$ global symmetry has been interpreted as coming
from the VEV of the auxiliary component field of the radion
superfield, and thus being interpreted as spontaneous breaking of
supersymmetry, in a five dimensional (5D) theory compactified on an
orbifold~\cite{luty,marti,kaplan}. Finally supersymmetry-breaking
induced by brane localized dynamics has been studied in
several recent papers~\cite{nilles,fabio2,tony}.

In this paper we want to study the nature of the symmetry that protects
the softness of the Scherk-Schwarz breaking in higher dimensional theories and
the finiteness of the Higgs mass after electroweak symmetry breaking.
We consider 5D theories compactified on the orbifold
$S^1/\mathbb{Z}_2$, which allows for chiral matter in the massless
sector and thus is suitable for phenomenological applications.  We
prove that local supersymmetry spontaneously broken by Wilson lines is
equivalent to SS supersymmetry breaking. In fact the $SU(2)_R$
symmetry is gauged in off-shell $N=2$ supergravity and 5D $SU(2)_R$
gauge bosons are auxiliary fields in the minimal supergravity
multiplet. When the corresponding Wilson lines acquire a VEV,
supersymmetry is spontaneously broken and one recovers the
SS-mechanism where different boundary conditions are imposed to
non-trivial representations of $SU(2)_R$. The would-be Goldstinos are
absorbed by the corresponding gravitinos in the unitary gauge
(super-Higgs mechanism~\cite{cremmer}) and disappear from the physical
spectrum of fields. In summary, local $N=2$ supersymmetry is
spontaneously broken by Wilson lines. The breaking survives in the
flat limit, where we decouple all gravitational interactions, and the
theory becomes softly broken global supersymmetry. No new counterterms
are created by the spontaneous breaking of the gauge (and super)
symmetry, which explains the finiteness of the Higgs mass under
radiative corrections provided the theory does not contain anomalous
$U(1)$ gauge groups~\footnote{In the presence of a $U(1)$ gauge
symmetry one has to take into account a possible Fayet-Iliopoulos (FI) 
term on the
brane which would contribute to the value of the Higgs mass. Since this is
not an F-term it is not protected by the
supersymmetric non-renormalization theorems but only by the vanishing of the
trace of the $U(1)$ charge matrix $Y$. In theories with $\text{tr}~Y\neq
0$ the FI-term is quadratically sensitive to the UV cutoff~\cite{fi} and thus
must be fine-tuned in order to give a small Higgs mass. In theories
with $\text{tr}~Y=0$ it is not renormalized and can be
consistently set to zero.}.


\section{\sc Wilson lines vs.~Scherk-Schwarz breaking}
In theories with extra dimensions compactified on a torus or an
orbifold, a symmetry can be broken by two mechanisms which are not
present in simply connected spaces: the Scherk-Schwarz and the
Hosotani (Wilson line) mechanisms.  In the Scherk-Schwarz
mechanism~\cite{SS,SSstring}, fields possess non-trivial boundary
conditions and are multiple-valued along the extra dimension. Let us
assume that a global or local symmetry with generator $\mathcal Q$ is
used to twist some of the fields under $2\pi R$-translations:
\begin{equation}        
\phi(x^\mu,\, x^5+2\pi R)=
e^{2 \pi i \omega \mathcal Q}\phi(x^\mu,\, x^5)\ .
\label{boundary}
\end{equation}
This periodicity condition is satisfied if the fields are given
by~\footnote{It has been pointed out~\cite{fabio1,fabio2} in the
context of brane-induced supersymmetry breaking that there exist a more
general class of solutions to Eq.~(\ref{boundary}).}
\begin{equation}
\phi(x^\mu,\, x^5)=
e^{ i \omega \mathcal Q\, x^5/R}\tilde\phi(x^\mu,\,x^5 )\ ,
\label{ydep}
\end{equation}
where $\tilde\phi(x^\mu,\, x^5+2 \pi R)=\tilde\phi(x^\mu,\, x^5)$. 
The symmetry generated by $\mathcal Q$ is then broken at tree level by the
kinetic term. 

If the symmetry with $\mathcal Q$-generator
is a local one it is possible to break it by giving a
VEV to an extra component of the corresponding gauge 
field $A_M$ ($M=\mu,5$)~\cite{hosotani} 
(Hosotani or Wilson line mechanism). All fields are
periodic in this picture and gauge transformations which preserve this
periodicity (the ones with periodic parameters) divide the
Wilson lines into equivalence classes which represent the possible
vacua of the theory. We can easily label these vacua by the constant
configuration of each equivalence class. We will assume from now on a
constant $\left<A_5\right>$.
All non-singlet fields will receive a mass-shift relative to their
KK-value through their covariant derivative. For instance, for a flat
compact direction the mass of KK-modes $\phi^{(n)}$ is given by
\begin{equation}
\mathcal M_{\phi^{(n)}}=\frac{n}{R}~{\bf 1}_G+\langle A_5\rangle,
\end{equation} 
with ${\bf 1}_G$ the identity in the fundamental representation of
the gauge group G (for non-constant VEVs the mass matrix will not be
block-diagonal). It is possible to switch to the Scherk-Schwarz
picture by allowing for gauge transformations with non-periodic
parameters. By choosing a transformation as
\begin{equation}
\exp\left\{i\,\left<A_5\right>\, x^5\right\}
\end{equation}
to transform away the VEV we end up with non-periodic fields as in
Eq.~(\ref{boundary}), with $\omega \mathcal Q\equiv R\left<A_5\right>$.  

\section{\sc Five-dimensional $N=2$ offshell supergravity}
To achieve supersymmetry-breaking one usually applies the SS-mechanism
to the automorphism group $SU(2)_R$ of the $N=2$ supersymmetry in five
dimensions. It would be desirable to have the Hosotani picture in this
case as well. The softness of the SS-breaking (i.e.~no additional
counterterms except the ones allowed by unbroken supersymmetry) could
then easily be explained by the fact that Hosotani breaking is
spontaneous. Steps in this direction have been recently
taken~\cite{marti,kaplan} by giving a VEV to the $F$-term of the
radion field: when coupling the radion superfield to matter charged
under $SU(2)_R$ the SS spectrum is reproduced.  However the final
proof that such a breaking is related to the Hosotani mechanism, and
therefore equivalent to SS, requires the introduction of the
gravitational sector.

To this end we will consider the recently found off-shell version of
$5D$ $N=2$ supergravity~\cite{zucker,kugo}.  The complete off-shell
particle content together with the orbifold-parity assignments are
displayed in table \ref{minimal}.
\begin{table}[t]
\begin{center}
\begin{tabular}{||cccccc||}
\hline\hline
field & description &dim& states&even&odd\\
\hline
$g_{MN}$ & graviton &0&
        $(15-5)_ B$&$g_{\mu\nu}$, $g_{55}$&$g_{\mu 5}$\\
$\psi_M$&gravitino&2&$(40-8)_F$& 
        $\psi_{\mu L}^1$, $\psi_{5L}^2$&$\psi_{\mu L}^2$, 
	$\psi_{5 L}^1$\\
$B_M$&graviphoton&3/2&$(5-1)_B$&
        $B_5$&$B_\mu$\\
$\vec V_M$&$SU(2)_R$ gauge field&5/2&
        $(15-3)_B$&$V_\mu^3$, $V_5^{1,2}$&$V_\mu^{1,2}$, $V_5^3$\\
$v^{AB}$&antisymmetric tensor&5/2&$10_B$&$v^{\alpha 5}$&
	$v^{\alpha \beta}$\\
$\vec t$&$SU(2)_R$ triplet&5/2&$3_B$&$t^{1,2}$&$t^3$\\
$C$&real scalar&7/2&$1_B$&$C$&\\
$\zeta$&$SU(2)_R$ spinor&3&$8_F$&$\zeta_L^1$&$\zeta_L^2$\\
\hline\hline
\end{tabular}
\end{center}
\caption{The field content of the minimal supergravity multiplet in
five dimensions which has $40_B+40_F$ components.}
\label{minimal}
\end{table}
Choices for parity assignments are discussed in Ref.~\cite{parities}.
 They have to be consistent with the symmetries of the bulk action and
 are thus mainly dictated by $N=2$ supersymmetry and 5D Poincar\'e
 invariance. All fields but $g_{MN},\psi_M$ and $A_M$ are auxiliary
 only.  The fermions are of the symplectic Majorana type and carry an
 $SU(2)_R$ index. They can be decomposed into two component spinors as
\begin{equation}
\label{simM}
\lambda^i = \left(
\begin{matrix}
\lambda_L^i\\\epsilon^{ij}\bar\lambda_{Lj}
\end{matrix}
\right)
\end{equation}
where we use the convention $\epsilon^{12}=+1$. Frequently we will 
suppress the $SU(2)_R$ index. 
The standard procedure to construct the theory is covariantizing the
globally supersymmetric transformation rules with respect to local
Lorentz and local $SU(2)_R$ transformations and then
supercovariantizing with respect to local supersymmetric
transformations.  After the total covariantization the supersymmetric
transformations include covariant derivatives of component fields and
of the supersymmetric parameter $\xi^i(x^M)$, as well as additional
terms of $\mathcal{O}(\kappa)$ and higher to close the 
algebra~\footnote{We denote by $\kappa$ the five dimensional gravitational
coupling which has mass dimension $-3/2$.}. The
complete expression for these transformations have been given in
Ref.~\cite{zucker}.  It will be particularly relevant for our
purposes the transformation of the gravitino $\psi_M$ which 
contains the covariant derivative of the transformation parameter:
\begin{equation}
\label{gravitino}
\delta_\xi\psi_M=\frac{1}{\kappa}\mathcal{D}_M\,\xi+\dots
\end{equation}
Here $\mathcal{D}_M$ is the covariant derivative with respect to local
Lorentz and local $SU(2)_R$ transformations
\begin{equation}
\label{der}
\mathcal{D}_M=D_M+i\kappa \vec V_M\vec\tau\ ,
\end{equation}
while $D_M$ is the covariant derivative with respect to local Lorentz
transformations. The auxiliary field $V_M$ has mass dimension $5/2$
as shown in table~\ref{minimal}.
Notice that the local supersymmetric transformations,
e.g.~in Eq.~(\ref{gravitino}), are consistent with the orbifold action
if we define $\xi^1_L(x^M)$ as an even function and $\xi^2_L(x^M)$ as
an odd one.  Together with the parities of the covariant derivatives,
i.e.~$\mathcal{D}_\mu$ even and $\mathcal{D}_5$ odd, this gives equal
parities on both sides of Eq.~(\ref{gravitino}).

Moreover it has been emphasized that the minimal multiplet is too
small to construct a physical action out of it~\cite{zucker}. It has
to be extended by a matter multiplet with $8_B+8_F$ components to
complete the $48_B+48_F$ components that a supergravity multiplet must
have. One possibility is adding the non-linear multiplet with field
content: $\phi^i_\alpha$, a scalar where the index $\alpha$ transforms
as a doublet of a global $SU(2)$; $\varphi$ a singlet scalar; $V_M$, a
singlet vector, and; $\chi^i$, an $SU(2)_R$ doublet fermion. The
supersymmetric transformations for the non-linear multiplet components
were given in Ref.~\cite{zucker} where it is shown that closure of the
algebra on $\varphi$ and $V_M$ requires a supersymmetric constraint
between the components of the minimal and non-linear multiplets which
removes the extra bosonic component, with respect to the fermionic
ones, which appears in the non-linear multiplet. After doing that the
kinetic Lagrangian for $SU(2)_R$ doublets, i.e.~the 5D gravitino
$\psi_M^i$ in the minimal supergravity multiplet, gauginos $\lambda^i$
in 5D vector multiplets and scalars $A^i$ in hypermultiplets can be
written as
\begin{equation}
\label{lagrange}
\mathcal{L}_{\,\text{kin}}=
-\frac{i}{2}\bar\psi_M \gamma^{MNP}\mathcal{D}_N\psi_P
+\frac{i}{2}\text{Tr}\,\left( \bar\lambda\gamma^M \mathcal{D}_M \lambda\right)
-\text{Tr} \left |\mathcal{D}_M A\right |^2\ ,
\end{equation}
where the covariant derivatives are given in (\ref{der}). Full actions 
for pure supergravity and supergravity coupled to $N=2$
matter multiplets can be found in Refs.~\cite{zucker,kugo}. 

\section{\sc Wilson breaking of local supersymmentry}
To perform a Wilson line breaking we have to consider supergravity
since local $SU(2)_R$ requires local supersymmetry.  However we will
see that at the end one can consistently decouple all gravity
interactions by taking the limit $\kappa\rightarrow 0$ while retaining
the effect of supersymmetry breaking. In the minimal realization of 5D
$N=2$ supergravity described above the $SU(2)_R$ symmetry becomes
local but it is gauged only off-shell, i.e. the
corresponding gauge fields $\vec V_M$ are auxiliary fields. Upon
orbifolding some of the gauge bosons are even, $V_\mu^3$, and the
others, $V_\mu^{1,2}$, odd and so happens with the corresponding gauge
transformations,
\begin{equation}
U=\exp (i\, \vec\Lambda\vec\tau)\ ,
\end{equation}
where $\Lambda^3$ is even and $\Lambda^{1,2}$ odd. This guarantees that the
$\mathbb{Z}_2$ parity defining the orbifold in table~\ref{minimal}
is stable under gauge transformations. 
In this environment one can use the triplet
of scalars $\vec{V}_5$ to break $SU(2)_R$ by Wilson lines.
As presented in table~\ref{minimal},
the fifth components of the gauge bosons $\vec{V}_5$ should have opposite
parities: $V_5^3$ is odd and $V_5^{1,2}$ are even. These assignments are also
consistent with the requirement that the supercovariant derivatives 
(\ref{der}), which include local $SU(2)_R$ transformations
have a well defined parity. Notice that the parity operator acting on
$SU(2)_R$ doublets is proportional to $\tau^3$, which anticommutes
with $\tau^{1,2}$. As a consequence $\tau^{1,2}$ in (\ref{der})
make, when acting on doublets, a parity flip required for consistency
of supercovariant derivatives. In particular, there is contribution
from the even scalars $V_5^{1,2}$ to the odd derivative $\mathcal D_5$
in (\ref{der}). 

Now we would like to turn a (constant) VEV on $\vec V_5$ in order to
get nontrivial Wilson lines. Observe that by five dimensional gauge
invariance there is no potential for $\vec V_5$ which is thus a flat
direction. Our choice for the parity of the fields allows constant
VEVs only for $V_5^1$ and $V_5^2$. Without loss of generality we can
choose $V_5^2$ and require
\begin{equation}
\left<V_5^2\right>=\frac{\omega}{\kappa R}.
\label{vev}
\end{equation}
This parametrization gives the correct dimension and also the right
behaviour in the decompactification and flat limits.
The factor $1/R$ in (\ref{vev}) guarantees that the VEV vanishes in
the limit $R\to\infty$, where the compact dimension becomes infinite,
since the corresponding Wilson breaking should disappear in the limit
of ``non-compact'' (infinite) extra dimension.
On the other hand, sending $\kappa\rightarrow 0$ we will see that the
gravitino mass remains constant as desired.
The dimensionless factor $\omega$ is identified with the SS-parameter.

The Goldstino is found from the local 
supersymmetric transformations and it is provided
by the fifth component of the gravitino, $\psi_5$, which has 
a nonzero transformation in the vacuum:
\begin{equation}
\delta_\xi \psi_5=\frac{1}{\kappa}{\cal D}_5\xi\ .
\label{goldsloc}
\end{equation} 
Let us analyze the corresponding super-Higgs~\cite{cremmer}
effect.
The kinetic terms for the gravitino can be
decomposed in four-dimensional and extra components as:
\begin{eqnarray}
-\frac{i}{2}\bar\psi_M\gamma^{MNR}{\cal D}_N\psi_R&=&-\frac{i}{2}
\bar\psi_\mu\gamma^{\mu\nu\rho}{\cal D}_\nu\psi_\rho
+\frac{i}{2}\bar\psi_\mu\gamma^{\mu\nu}\gamma^5{\cal D}_5\psi_\nu \nonumber\\
&&-\frac{i}{2}\bar\psi_\mu\gamma^{\mu\nu}\gamma^5{\cal D}_\nu\psi_5
-\frac{i}{2}\bar\psi_5\gamma^{\mu\nu}\gamma^5{\cal D}_\mu\psi_\nu\ .
\end{eqnarray}
We now do the redefinition~\footnote{For constant non-zero $\omega$
the operator ${\cal D}_5$ is non-singular.}
\begin{equation}
\psi_\mu=\psi_\mu'+{\cal D}_\mu\left({\cal D}_5\right)^{-1}\psi_5\ ,
\label{unitary}
\end{equation}
which can be seen as a local supersymmetry-transformation with
parameter $\kappa({\cal D}_5)^{-1}\psi_5$ gauging $\psi_5$ away.  
We arrive at a ``unitary gauge'' where $\psi_5$ has been ``eaten''
by the 4D gravitino $\psi_\mu$:
\begin{equation}
-\frac{i}{2}\bar\psi_M\gamma^{MNR}{\cal D}_N\psi_R=
-\frac{i}{2}\epsilon^{\mu\nu\rho\sigma}
\bar\psi'_\mu\gamma_\sigma\gamma^5{\cal D}_\nu\psi'_\rho
+\frac{i}{2}\bar\psi'_\mu\gamma^{\mu\nu}\gamma^5  
 \left(\partial_5+i\frac{\omega}{R}\tau^2+...\right)\psi'_\nu\ .
\end{equation}
The second term provides the mass term for $\psi'_\mu$,
which can be expressed in two component
spinors (now omitting any $'$):
\begin{equation}
\mathcal L_{\rm mass}=\frac{1}{2}
\left(\psi^1_{\mu L}\ \psi^2_{\mu L}\right)
\sigma^{\mu\nu}
\left(
\begin{array}{cc}
\omega/R & i\,p_5\\
-i\,p_5 &\omega/R
\end{array}
\right)\left(
\begin{array}{c}
\psi^1_{\nu L} \\
\psi^2_{\nu L}
\end{array}\right)+h.c.
\label{masagrav}
\end{equation}
Here $p_5\equiv i\,\partial_5$ is the component of the momentum along
the extra dimension and -- for a flat extra dimension -- takes the
values $n/R$, where $n$ is a non-negative integer. We can see that the
flat direction $\left<V_5^2\right>$ breaks supersymmetry as in the
no-scale models~\cite{noscale} where the gravitino zero mode mass is
$m_{3/2}=\omega/R$.  

Note that Eq.~(\ref{masagrav}) holds even for
$x_5$-dependent $\omega$ provided that the transformation (\ref{unitary})
is nonsingular. In particular, making the extreme choice
\begin{equation}
\omega \propto \delta(x_5)P_0+\delta(x_5-\pi R)P_\pi\ ,
\label{brane}
\end{equation}
where $P_0$ and $P_\pi$ are the constants parametrizing superpotential
VEVs in Ref.~\cite{fabio2}, we can also incorporate in our off-shell
formulation brane-induced supersymmetry breaking and Eqs.~(9) and (11)
of~\cite{fabio2} are reproduced. The main task would then be to
diagonalize the mass matrix as done there. The case $P_0=-P_\pi$
results in trivial Wilson lines and supersymmetry remains unbroken. A
possible physical meaning for the choice (\ref{brane}) was given in
Refs.~\cite{nilles,fabio2,tony} where it was interpreted as the result
of some integrated out brane fields leaving a constant superpotential
on each brane~\footnote{Terms proportional to
$\delta(x_5-x_5^{\rm brane})$ can generally be created by loop effects
of bulk and brane couplings~\cite{braneff}. Consideration of loop
corrections is beyond the scope of the present paper. }.

The described formalism is of course equally well applicable if one
compactifies on $S^1$ instead of $S^1/{\mathbb Z}_2$. Five-dimensional
$N=2$ local supersymmetry is directly broken down to $N=0$ by giving
a VEV to $V_5^i$.

Writing the kinetic terms for the gauginos and the hyperscalars
explicitly we get
\begin{eqnarray}
\frac{i}{2}\,\bar\lambda\gamma^M{\cal D}_M\lambda&=&
\frac{i}{2}\,\bar\lambda\gamma^\mu{\cal D}_\mu\lambda
        +\frac{i}{2}\,\bar\lambda\gamma^5\left(\partial_5+
i\frac{\omega}{R}\tau^2+...\right)\lambda\ ,\nonumber\\
A^\dagger{\cal D}^2A&=&A^\dagger{\cal D}^\mu{\cal D}_\mu A
        -A^\dagger\left(\partial_5+i\frac{\omega}{R}\tau^2+\dots\right)^2A\ ,
\label{derexp}
\end{eqnarray}
while the corresponding superpartners are inert under $SU(2)_R$.
The second terms provide the shifted masses for gauginos and hyperscalars.
For gauginos the mass Lagrangian can be cast as
\begin{equation}
\mathcal L_{\rm mass}=
\frac{1}{2}\left(\lambda^1_L\ \lambda^2_L\right)
\left(
\begin{array}{cc}
\omega/R & i\,p_5\\
-i\,p_5 &\omega/R
\end{array}
\right)\left(
\begin{array}{c}
\lambda^1_L \\
\lambda^2_L
\end{array}\right)+h.c.
\label{masag}
\end{equation}
Similarly for the hyperscalars one obtains,
\begin{equation}
\mathcal L_{\rm mass}=-
\left(A_1^\dagger\ A_2^\dagger\right)
\left(
\begin{array}{cc}
p_5^2+(\omega/R)^2 & i\,2\omega\,p_5/R\\
-i\,2\omega\,p_5/R & p_5^2+(\omega/R)^2
\end{array}
\right)\left(
\begin{array}{c}
A^1 \\
A^2
\end{array}\right)\ .
\label{masah}
\end{equation}
Mass matrices for (\ref{masagrav}), (\ref{masag}) and (\ref{masah})
give rise to the mass eigenvalues $\omega/R\pm p_5$, identical to
those obtained in SS-broken theories \cite{matter,UV}.

It is easy to see looking at Eq.~(\ref{derexp}) that we still have
supersymmetry breaking in the global limit $\kappa\rightarrow 0$,
$m_{3/2}$ fixed.  Taking this limit we decouple all gravity degrees of
freedom and are left with $N=1$ global supersymmetry which is broken
down to $N=0$ whenever we have nonzero $\omega$.  In the case of a
circle, going to the flat limit results in $N=2$ global supersymmetry
broken down to $N=0$.  One can also see some connection of the
Hosotani/SS-mechanism with the radion mediation of
Refs.~\cite{marti,kaplan}. By looking at the linearized (global)
transformations one finds that the radion multiplet~\footnote{$h_{MN}$
denotes the linearized metric $g_{MN}=\eta_{MN}+\kappa h_{MN}$.}
\[
\left(h_{55}+iB_5,\ \psi_{5L}^2,\ V_5^1+i V_5^2+4i(t^1+it^2)\right)
\]
transforms separately under global supersymmetry.  The auxiliary field
of the radion contains explicitly $V_5^1+iV_5^2$, so the $F$-term of
this superfield will get a VEV as well. In the local theory it induces
spontaneous breaking of supersymmetry, but in the global limit the
would-be Goldstino is decoupled and the breaking of supersymmetry is soft.
Finally the radion VEV remains
undetermined at this level by the no-scale nature of the
SS-breaking.

\section{\sc Conclusions}
We conclude that supersymmetry breaking is mediated already at tree
level from the gravity to the matter/gauge sector through the local
$SU(2)_R$ couplings.  
In complete analogy to the case of an ordinary gauge symmetry, the
equivalence to SS breaking can be seen by transforming away the VEV of
$V_5^2$ using a multivalued gauge transformation.
Finally the softness of the SS-mechanism is explained by
the fact that Hosotani breaking is spontaneous. Only counterterms
which are allowed by the underlying supersymmetry can appear in the
Lagrangian and so the usual non-renormalization theorems apply. This
confirms recent results found in explicit one and two loop
calculations~\cite{one,one1,one2}.

We summarize the supersymmetry breaking scheme in the following
diagram. We denote by $(N=0)_{\text{local, global}}$ the 
breakdown of the corresponding $(N\neq0)_{\text{local, global}}$.  
\vspace{.25cm}

\noindent
\fbox{\parbox{15.5 cm}
{\footnotesize
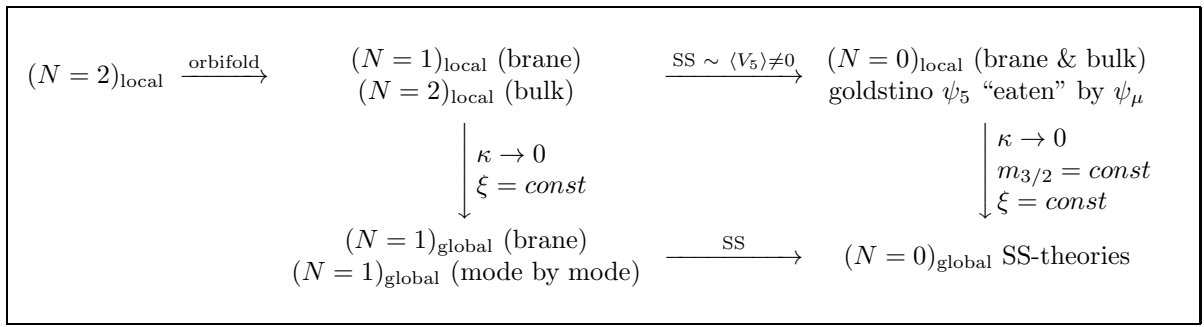
\[
\arraycolsep0.1cm
\begin{array}{ccccc}
\hspace{-.2cm}(N=2)_{\text{local}}  & \xrightarrow{\rm orbifold}
&\begin{array}{c}(N=1)_{\text{local}}\text{ (brane)}\\
(N=2)_{\text{local}}\text{ (bulk)}\end{array}&
\xrightarrow{\text{SS }\sim\ \left<V_5\right>\neq0 }
&\begin{array}{c}(N=0)_{\text{local}}\text{ (brane \& bulk)}\\
\text{goldstino }\psi_5 
\text{ ``eaten'' by }\psi_\mu \end{array}\vspace{.2cm}\\
&&
\left\downarrow\vcenter{\rlap{
\!\!\!$\begin{array}{l}\kappa\to 0\\\xi=const\end{array}$	
$\begin{array}{l}\ \\\ \\\	 \end{array}$}}\right.
&&
\left\downarrow\vcenter{\rlap{
\!\!\!$\begin{array}{l}\kappa\to 0\\m_{3/2}=const\\\xi=const\end{array}$
}}\right .\vspace{.1cm}\\
&&\begin{array}{c}(N=1)_{\text{global}}\text{ (brane)}\\
(N=1)_{\text{global}}\text{ (mode by mode)}
\end{array}&
\xrightarrow{\hspace{0.65cm}\text{SS}\hspace{.65cm}}&\begin{array}{c}
(N=0)_{\text{global}}\text{ SS-theories}
\end{array}
\end{array}
\]
}
}
\vspace{.25cm}

Starting from a genuine $5D$ $N=2$ locally supersymmetric theory we
compactify on the orbifold $S^1/\mathbb{Z}_2$ thus creating two branes
with local $N=1$ supersymmetry. In the bulk we retain $N=2$ local
supersymmetry with a $\mathbb{Z}_2$ constraint on the parameter $\xi$,
i.e.~$\xi_L^1$ is even and $\xi_L^2$ odd. Giving a VEV to $V_5^2$
breaks both supersymmetries spontaneously, the corresponding Goldstino
$\psi_5$ providing the longitudinal components of the 4D gravitino
$\psi_\mu$. This mechanism is equivalent to SS-breaking by means of a
nonperiodic $SU(2)_R$ transformation. Taking $\kappa\rightarrow 0$ one
arrives then at the SS-broken theories widely considered in the
literature~\cite{matter,UV}.  If SS/Wilson breaking is absent, one
obtains in the flat limit $N=1$ supersymmetry for the KK-modes (left
vertical arrow). The counterintuitive reduction to $N=1$ is explained
by the fact that the odd parameter $\xi_L^2$ vanishes in the global
limit, while the
linearity of the global transformation laws ensures that supersymmetry
is realized mode by mode~\cite{AHGW,progress}.


\section*{\sc Acknowledgments} 
We thank A.~Delgado for useful discussions and for participating at
the early stages of this work. The work of GG was supported by the
DAAD.

\end{document}